# Electronic *g*-factor and Magneto-transport in InSb Quantum Wells

Zijin Lei[1*], Christian A. Lehner[1], Km Rubi[2], Erik Cheah[1], Matija Karalic[1], Christopher Mittag[1],

Luca Alt[1], Jan Scharnetzky[1], Peter Märki[1], Uli Zeitler[2], Werner Wegscheider[1],

Thomas Ihn[1], and Klaus Ensslin[1]

1. *Solid State Physics Laboratory, ETH Zurich, CH-8093 Zurich, Switzerland*

2. *High Field Magnet Laboratory (HFML-EFML), Radboud University, 6525 ED Nijmegen, The Netherlands*

[*]*Email of the corresponding author: zilei@phys.ethz.ch*

**Abstract:**

High mobility InSb quantum wells with tunable carrier densities are investigated by transport experiments in magnetic fields tilted with respect to the sample normal. We employ the coincidence method and the temperature dependence of the Shubnikov-de Haas oscillations and find a value for the effective *g*-factor of $|g^*| = 35 \pm 4$ and a value for the effective mass of $m^* \approx 0.017\, m_0$, where $m_0$ is the electron mass in vacuum. Our measurements are performed in a magnetic field and a density range where the enhancement mechanism of the effective *g*-factor can be neglected. Accordingly, the obtained effective *g*-factor and the effective mass can be quantitatively explained in a single particle picture. Additionally, we explore the magneto-transport up to magnetic fields of 35 T and do not find features related to the fractional quantum Hall effect.

---

The narrow-gap III-V binary compound InSb is well known for its combination of a light effective mass, high electron mobility, strong spin-orbit interactions, and a giant effective *g*-factor in the conduction band [1-6]. These unique properties are interesting in a view of potential applications such as high-frequency electronics [7], optoelectronics [8], and spintronics [9]. Especially, it has been suggested that the large effective *g*-factor of InSb with a bulk value of |*g**| ~ 51 could be advantageous for hosting a topologically nontrivial phase through proximity-induced superconductivity [10, 11]. The effective *g*-factor has led to research efforts in nanomaterials and sophisticated nanoconstrictions. In 2-dimensional systems, so-called coincidence measurements have been used in InSb quantum wells (QWs). Here the magnetic field is tilted with respect to the sample surface normal to tune the relative strength of Zeeman and Landau level splitting [12, 13]. The effective *g*-factor has been measured in nanoconstrictions too, such as in nanowire-based quantum dots [14, 15], and in QW-based quantum point contacts [16, 17].

Nevertheless, all the experimental results introduced above focused on the situations of special quantum regimes or with relatively low Landau level filling factors where exchange effects dominate. These lead to an effective *g*-factor far off from the predictions of $\boldsymbol{k}\cdot\boldsymbol{p}$ theory. For example, in InSb nanowire-based quantum dots, the effective *g*-factor is more than 40% larger than that in the bulk material because of the level-to-level fluctuations arising from spin−orbit interaction [14, 18]. Caused by the confinement of the nanoconstrictions, the effective *g*-factor measured in chemically-etch defined quantum point contacts (QPCs) shows strong reductions and large anisotropies [16, 17]. Limited by the low carrier mobilities, the coincidence measurement in a tilted magnetic field can only be accomplished at a relatively low filling factor. The large spin polarization leads to a substantial enhancement of the *g*-factor as a result of exchange interactions [13, 14, 19]. Different from the results listed above, the bare *g*-factor deserves to be investigated for the research related to topological superconductivity, because the bandgap should be opened by the Zeeman splitting while applying a relatively small parallel magnetic field [20]. Recent measurement results from purely gate-defined quantum dots fabricated on InSb QWs present the estimation of the effective *g*-factor of a value between 26 and 35 [21]. This is close to the bare effective *g*-factor resulting from the $\boldsymbol{k}\cdot\boldsymbol{p}$ theory. However, the accuracy is still limited by the quality of the devices. In the experiment presented in this manuscript, we show a precise measurement of the bare *g*-factor by performing a coincidence measurement for InSb two-dimensional electron gas (2DEG) confined between InAlSb barriers. The high electron mobilities (up to $3\times 10^5$ cm$^2$/Vs) enable us to investigate the dependence of the Shubnikov-de Haas (SdH) oscillations on the tilt angle of the magnetic field relative to the plane of the 2DEG in a relatively small total magnetic field and for large filling factors. With the combination of a precise measurement of the effective mass, the effective *g*-factor of the InSb 2DEG turns out to be $|g^*| = 35 \pm 4$, which can be quantitatively explained with a single particle calculation based on $\boldsymbol{k} \cdot \boldsymbol{p}$ theory. Finally, we present electron transport in magnetic fields as high as 35 T. We find no signatures related to the fractional quantum Hall effect which asks for further improvement of the sample quality in the future.

The InSb QW sample we investigate here is grown on a (100) GaAs substrate by molecular beam epitaxy (MBE). A schematic layer sequence is shown in Fig. 1 (a). A specialized interfacial misfit transition to a GaSb buffer and an interlayer InAlSb buffer is employed to overcome the lattice mismatch between GaAs and InSb. The total thickness of the buffer system amounts to roughly 3 $\mu$m. Then, the 21 nm-thin InSb quantum well is surrounded by In$_{0.9}$Al$_{0.1}$Sb confinement barriers, while the n-type carriers are introduced to the active region by two Si δ-doping layers incorporated 40 nm below and above the QW in the barrier, respectively. On the top of the QW, an In$_{0.9}$Al$_{0.1}$Sb layer with a thickness of 180 nm is grown. Finally, a 3 nm thick InSb capping layer is employed on the top of the sample to prevent the possible oxidization. More details about the MBE growth can be seen in Ref. 22.

The micro-fabrication process is similar to our previous work [23]. A standard Hall bar sample is defined using wet chemical etching with an etching depth of more than 270 nm, which is deeper than the Si δ-doping layer on the substrate side. Layers of Ge/Ni/Au evaporated after an Ar milling provide the Ohmic contacts. The sample is coated with a 40 nm thick aluminum oxide (ALO) dielectric layer using atomic layer deposition (ALD) at a temperature of 150 ° C. A high-temperature annealing step is unnecessary because the metal diffuses into the 2DEG during the heating during the ALD process. Finally, a Ti/Au top gate covering the Hall bar is deposited with electron beam evaporation.

Figure 1 (b) presents an optical image of the sample together with the schematic diagram of the magneto-transport measurement configuration. The part of the Hall bar we measured has a size of 310× 25 $\mu$m$^2$. The transport measurement is accomplished by using standard low frequency (77 Hz) lock-in techniques in a cryostat with a base temperature of 1.3 K and a rotatable magnetic field up to 8 T. Figure 1 (c) and (d) show the longitudinal and the transverse resistivities $\rho_{xx}$ and $\rho_{xy}$ where the top gate voltages $V_{TG}$ are 0 V and -1 V, respectively. Pronounced SdH oscillations in $\rho_{xx}$ and quantum Hall plateaus in $\rho_{xy}$ at $h/\nu e^2$ are visible, where ν is the filling factor. The inset in Fig. 1(c) shows that the Zeeman splitting starts from magnetic fields around 1.1 T. This measurement indicates that a single channel 2DEG system with a tunable density and a high mobility is embedded in the QW. Figure 1 (e) and (f) show the Landau fan diagrams of $\rho_{xx}$ and $\rho_{xy}$, i.e., the dependence of $\rho_{xx}$ and $\rho_{xy}$ on perpendicular magnetic field $B$ and $V_{TG}$. The slight bending of the Landau fan diagram at high $V_{TG}$ may be caused by the charging of the capping layer or the barrier layer above the QW. The carrier density $n$ is calculated through both the linear fitting of the Hall effect in a small magnetic field window (from -0.3 T to 0.5 T) and the $1/B$ periodicity of the SdH oscillations. As shown in Fig. 1 (g), the carrier density can be tuned from $2 \times 10^{11}$ cm$^{-2}$ to $3.5 \times 10^{11}$ cm$^{-2}$ while increasing the mobility from $1.3 \times 10^5$ cm$^2$/(Vs) to $3 \times 10^5$ cm$^2$/(Vs). The gate capacitance is estimated to be $C = 0.32$ mF/m$^2$ at the linear part of the $n$-$V_{TG}$ line, which is about 40% less than the value calculated from the plane-parallel capacitor model, which is similar to the results of our previous work [23].

In the following, we present our results of the coincidence measurement when $n = 3.45 \times 10^{11}$ cm$^{-2}$ at $V_{TG} = -0.1$ V. Similar strategies were used for QWs with large effective $g$-factors before, such as Si/SiGe [24] and InAs [25]. The inset of Fig. 2 shows the definition of the tilt angle $\theta$, where $\theta = 0$ corresponds to a magnetic field perpendicular to the sample surface and $\theta = 90°$ means an in-plane magnetic field. Figure 2 depicts the dependence of $\rho_{xx}$ on $B_\perp$, the perpendicular component of the magnetic field, for a continuous change in $\theta$ from 0° to 84°. The value of $\theta$ is calibrated with the slope of the Hall effect in a low magnetic field with a high accuracy. The thick lines in Fig. 2 show magneto-resistance traces at specific $\theta$ values where the SdH minima occur only at even integer filling factors ($r = 2$), at even and odd filling factors ($r = 1/2, 3/2$), or only at odd-integer filling factors ($r = 1$). The trace with a purely perpendicular magnetic field ($\theta = 0$, $r = 0$) is labeled with a thick line too.

The inset in Fig. 3 (b) describes the various coincidence situations which are characterized by the parameter $r$, which represents the ratio of the Zeeman and the cyclotron energies:

$$r = |g^*|\mu_B B_{tot}/\hbar\omega_c .$$

Here $\omega_c = eB_\perp/m^*$, $m^*$ is the effective electron mass, $\mu_B$ is the Bohr magneton, and $B_{tot}$ is the total magnetic field applied, with which we can achieve $B_\perp = \cos(\theta)B_{tot}$. Thus, we can calculate $r\cos(\theta) = |g^*|m^*/2m_0$, where $m_0$ is the free electron mass. At $r = 1/2$ and $3/2$, the minima of SdH traces will occur at both even and odd integer filling factors. For $r = 1$ or $2$, the minima occur at only odd or even integer filling factor. Similar to the SdH traces, the disappearance and the reappearance of the Hall plateaus at even and odd filling factors can be investigated as a function of increasing $\theta$. As shown in Fig. 3 (a), both even and odd filling factor plateaus can be seen when $r = 1/2$ and $3/2$, but only odd or even filling factor plateaus are observed when $r = 1$ or $2$.

One can determine the product of g-factor and effective mass by taking these results into consideration. Figure 3 (b) shows the relation between $1/r$ and the corresponding $\cos(\theta)$. The error bar is determined by the space of $\theta$ values between the traces with the angles of coincidence ($r = 1/2, 1, 3/2$ and $2$) and the neighboring traces which lose the coincidence. The accuracy of our measurement is mainly limited by the broadening of the Landau levels, especially when $r = 1/2$ and $r = 3/2$, where the SdH oscillations are not very pronounced in a low magnetic field regime. This linear relation indicates that $|g^*|m^*$ and therefore $|g^*|$ are constant within our measurement range, i.e., exchange enhancement effects are not relevant. The slope of the plot shows that the product of $|g^*|m^* = 0.60 \pm 0.07\ m_0$. Thus, only the effective mass is left to measure in order to extract the value of $|g^*|$.

In the following, we present the measurement of the effective mass in the same sample through the temperature dependence of the SdH oscillations in small magnetic fields. Figure 4 (a) shows the SdH oscillations for a carrier density $n = 3.44 \times 10^{11}$ cm$^{-2}$ measured at temperatures from 1.53 K to 15 K. The oscillating part of the resistivity $\Delta\rho_{xx}$ is obtained by subtracting the background of the magneto-resistance $\bar{\rho}_{xx}$. Figure 4 (b) presents the fit of the Dingle factor to $\ln(\Delta\rho_{xx}T_0/\bar{\rho}_{xx}T)$, where $T_0$ is the lowest temperature at which we measured the SdH oscillations [26]. The obtained effective mass is $m^* \approx 0.017\ m_0$ with an error smaller than $0.001\ m_0$ within the magnetic field range of the measurement (see the inset of Fig. 4 (c)). Figure 4 (c) shows the fitting to obtain the quantum lifetime using the effective mass discussed above. We extract a quantum lifetime $\tau_q = 0.10 \pm 0.01$ ps from the slope obtained from the linear fitting of the plotting of $1/B$ vs. $\ln(\frac{\Delta\rho_{xx}}{\bar{\rho}_{xx}}f(B,T))$, where $f(B,T) = (2\pi^2 k_B T/\hbar\omega_c)/\sinh(2\pi^2 k_B T/\hbar\omega_c)$. Considering the Drude scattering time $\tau_D = 2.87$ ps calculated from the Drude model at this carrier density, the Dingle ratio can be calculated as $\tau_D/\tau_q \approx 29$. This value, which is considerably larger than 1, indicates that our 2DEG is in the regime where the dominant scattering mechanism originates from long-range potential fluctuations [26-28]. On the other hand, we observe that $\mu \propto n$ from magneto-transport measurements. The linear relationship between $\mu$ and $n$ was also observed in a high-quality asymmetric InSb QW in Ref. 4. This implies that background impurity scattering is dominant [29] which leads to a short-range scattering potential. These two observations, the large Dingle ratio on the one-hand and the linear behavior of $\mu$ versus $n$ on the other hand, seem to lead to contradictory statements about the dominant scattering mechanism. We speculate that spin-orbit interactions which are relevant in InSb could play a role here, or some other phenomena which we do not understand yet. We would like to point out that a similar situation is found for InAs quantum wells, where an even larger Dingle ratio and a sublinear relationship between $\mu$ and $n$ are observed together [30].

Repeating the measurement for $n = 3.25 \times 10^{11}$ cm$^{-2}$ and $n = 2.9 \times 10^{11}$ cm$^{-2}$, we find the effective mass to be constant and the ratio $\tau_D/\tau_q$ is always ~ 29. Combining this with the coincidence measurement, we conclude that $|g^*| = 35 \pm 4$. Furthermore, we obtain the same results for another set of coincidence measurements at 90 mK with a lower carrier density ($n = 2.88 \times 10^{11}$ cm$^{-2}$).

Due to the high mobility of our sample, our coincidence measurement is accomplished in a regime where the exchange enhancement can be neglected. The change of the spin polarization $P$ is defined as $P = r/\nu$. In our measurement, the Landau levels in the InSb 2DEG are distinguishable from $r = 1/2$ to 2, and the tilt angle dependence of the SdH oscillations has been investigated from $\nu = 5$ to 28. These give a range of $P$ from 0.018 to 0.2, where there is no obvious $|g^*|$-$B_\perp$ dependence observed. Thus, we conclude from our estimation that the extracted value of $|g^*|$ is basically the bare effective $g$-factor of the 2DEGs in InSb QWs.

Our measurement results can be quantitatively compared to multi-band $\boldsymbol{k} \cdot \boldsymbol{p}$ theory where both the s- and p-like bands are considered [31]. Due to the confinement of the QW, the bandgap of InSb will have an increase of $E_g = \hbar^2/L^2 8\pi m^*_{\text{bulk}}$ based on the infinite potential well approximation, where $m^*_{\text{bulk}} \approx 0.014 \, m_0$ is the electron effective mass of bulk InSb and $L$ = 21 nm is the width of the QW. Thus, we can calculate as in Ref 26 and 31:

$$\frac{m_0}{m_0^*} = 1 + \frac{1}{3}\frac{2m_0 P^2}{\hbar^2}\left(\frac{2}{E_0+E_g} + \frac{1}{E_0+\Delta_0+E_g}\right)$$
$$- \frac{1}{3}\frac{2m_0 P'^2}{\hbar^2}\left(\frac{2}{E_0'-E_0+\Delta_0'} + \frac{1}{E_0'-E_0}\right), \quad (1)$$

and

$$g_0^* = 2 - \frac{2}{3}\frac{2m_0 P^2}{\hbar^2}\left(\frac{1}{E_0+E_g} - \frac{1}{E_0+E_g+\Delta_0}\right)$$
$$+ \frac{2}{3}\frac{2m_0 P'^2}{\hbar^2}\left(\frac{1}{E_0'-E_0} - \frac{1}{E_0'-E_0+\Delta_0'}\right), \quad (2)$$

where $m_0^*$ and $g_0^*$ are the theoretical electron effective mass and effective $g$-factor in the QW [26]. Here, we adopt the band edge parameters $E_0$ = 0.24 meV, $E_0'$ = 3.16 meV, $\Delta_0$ = 0.82 meV, $\Delta_0'$ = 0.33 meV, $P$=1.049 eVnm, and $P'$=0.481i eVnm from the Ref. 29. The calculation shows that the theoretical effective mass of the electrons in the QW has a value of $m_0^*$ = 0.017 $m_0$ and the effective $g$-factor has a value of $g_0^*$ = -37.2, which precisely agrees with our measurement results. We also find that there is a difference of 5% between the measurement and the calculation of the effective mass using a two-band ***k·p*** theory.

The high mobility of our sample motivates us to probe the behavior of the 2DEGs in an even higher magnetic field, where $v < 1$. We perform a magneto-transport measurement on another Hall bar sample with the same quality, where the heterostructure is grown on a GaSb wafer with the same growth and microfabrication. As shown in Fig. 5, measurements at 4 K and 0.4 K reveal similar $\rho_{xx}$ and $\rho_{xy}$ behaviors at $v < 1$, where the ohmic contacts still work properly. This implies that the 2DEG is not in an insulating phase. Despite the high quality of the sample, no fractional quantum Hall effect features are observable. The broad local minima of $\rho_{xx}$ at $B$ = 27 T and 33 T are far from the estimated magnetic field values corresponding to the filling factor values $v$ = 2/3 and 1/2 and their locations strongly depend on temperature. In addition, $\rho_{xy}$ at $v < 1$ increases above the classical limit, which is indicated by the dashed line in Fig. 5. Even though the InSb QW sample measured here exhibits comparably high mobility as shown in Ref. 22, the impact of the disorder or other effects on the transport behavior in the extreme quantum limit needs to be further investigated. This raises the question of whether material-specific properties are responsible for the observations since a similar behavior in InSb QWs has been observed before [32].

In conclusion, we present a measurement of the bare *g*-factor of an InSb QW with tunable density. Due to the high mobility, our measurement is accomplished in a low magnetic field regime where many-body physics can be neglected. Together with a precise measurement of the effective mass, the bare effective *g*-factor of our InSb QW is determined to be $|g^*| = 35 \pm 4$. Furthermore, probing the electron transport behavior for filling factors $\nu < 1$ is presented, where the FQHE is not observed.

We thank Dr. Folkert K. de Vries and Dr. Peter Rickhaus for fruitful discussions. We acknowledge the support of the HFML, member of the European Magnetic Field Laboratory (EMFL). This work was supported by the Swiss National Science Foundation through the National Center of Competence in Research (NCCR) Quantum Science and Technology.

FIG. 1. (a) Layer structure of the QW (b) An optical image of the Hall bar sample. The measurement setup is added schematically. (c) and (d) The $B$-dependence of $\rho_{xx}$ (blue) and $\rho_{xy}$ (red) for $V_{TG} = 0$ V and $V_{TG} = -1$ V are presented, respectively. The filling factors are labeled on the plateaus of the Hall traces or the minima of the SdH oscillations. Inset of (c): the zoom-in of (c) in the small magnetic field range. The axes are the same with (c) and (d). (e) and (f) The detailed transport characterization at 1.3 K with $\rho_{xx}$ (e) and $\rho_{xy}$ (f) as functions of $V_{TG}$ and $B$. The filling factors are labeled in white color on the Landau Fan diagrams. (g) The blue line and squares show the carrier density of the Hall bar obtained from the Hall effect $n_{Hall}$ and from the 1/$B$ periodicity of the SdH oscillations $n_{SdH}$ plotted against $V_{TG}$. The red line depicts the dependence of the electron mobility $\mu$ on $V_{TG}$.

FIG. 2 The SdH oscillations measured with different tilt angles. The traces have a constant offset of 40 Ω. The traces with $r$ = 0, 1/2, 1, and 2 are high lighted with thick lines, together with the corresponding angles. The integer filling factors are labeled with dashed lines. Inset: The definition of the tilt angle $\theta$.

FIG. 3 (a) The quantum Hall effect traces with $r$ = 0, 1/2, 1, and 2. The Hall plateaus are labeled with the dashed lines. (b) Coincidence plot gathered from the angles in Fig. 2. The straight line has a slope indicating $|g^*|m^* = 0.60 \pm 0.07\ m_0$. Inset: the various coincidence situations which are characterized by the parameter $r$, the ratio of the Zeeman and the cyclotron energies.

FIG. 4 Effective mass measurement. (a) Temperature-dependence of SdH oscillations with $n = 3.44 \times 10^{11} \text{cm}^{-2}$. (b) Dingle factor fitting with different $B$. The squares are data and the lines are fitted curves. (c) The fitting of the quantum lifetime. Inset: The effective mass obtained from the Dingle factor fitting vs. $B$.

FIG. 5 The $\rho_{xx}$ and $\rho_{xy}$ measurement of the InSb QW grown on a GaSb wafer in a large magnetic field range at the temperatures of 4 K and 0.4 K. The dashed line is an extrapolation of the Hall trace in the small magnetic field.

**Figure 1 (*This is a 2-column figure*)**

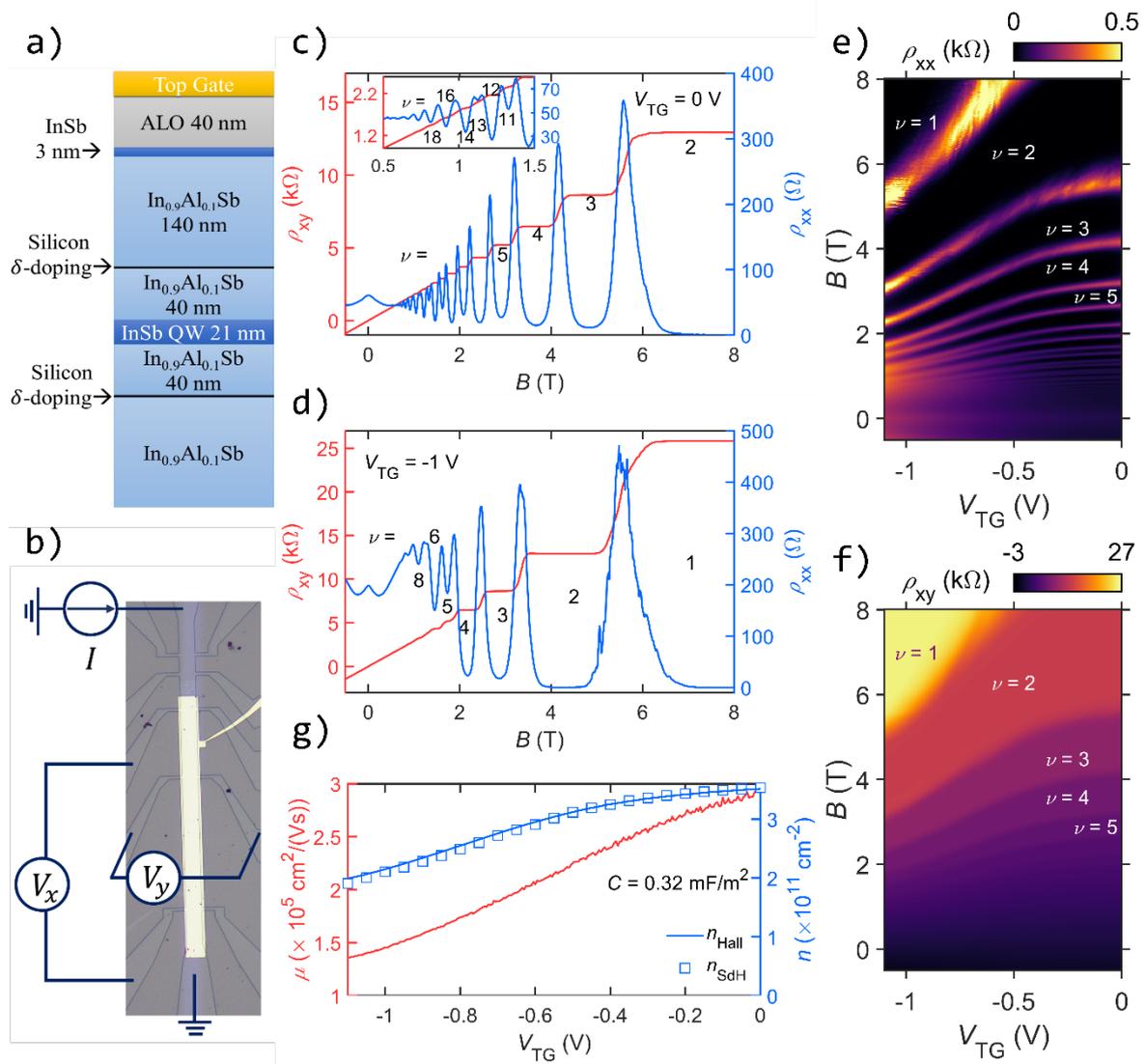

**Figure 2 (*This is a 2-column figure*)**

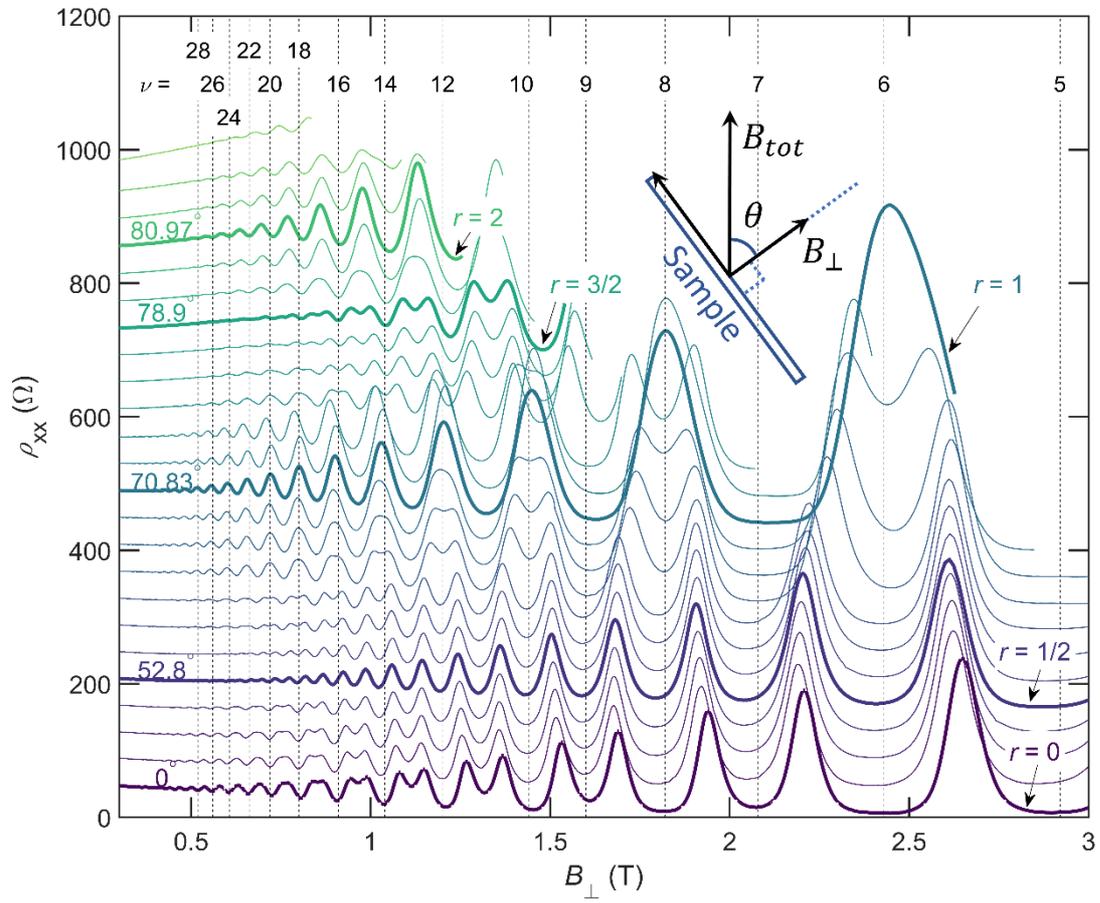

**Figure 3 (*This is a 1-column figure*)**

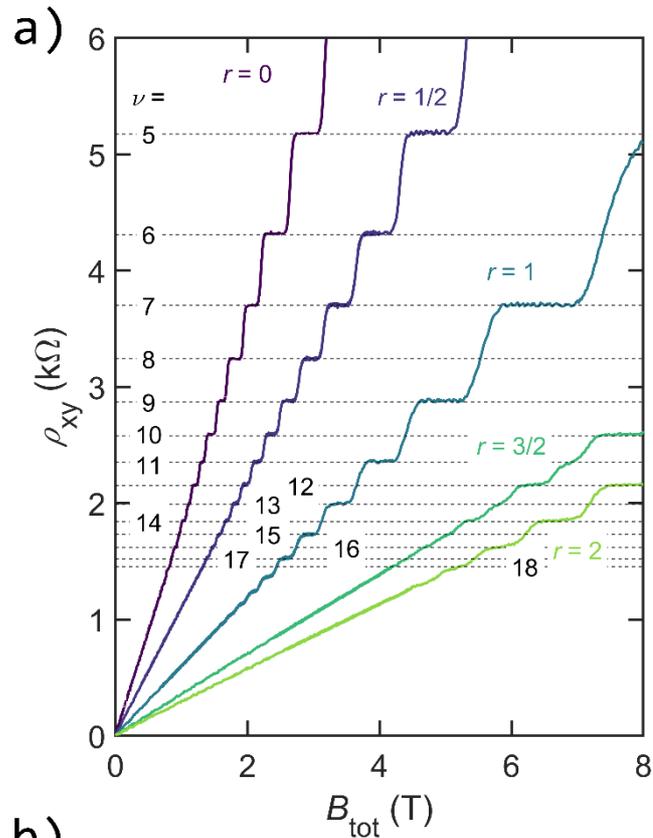

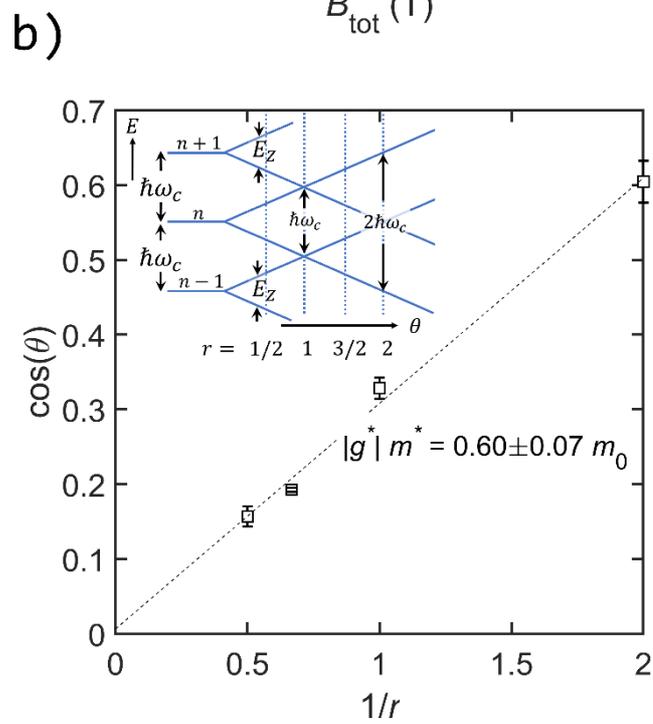

**Figure 4 (*This is a 1-column figure*)**

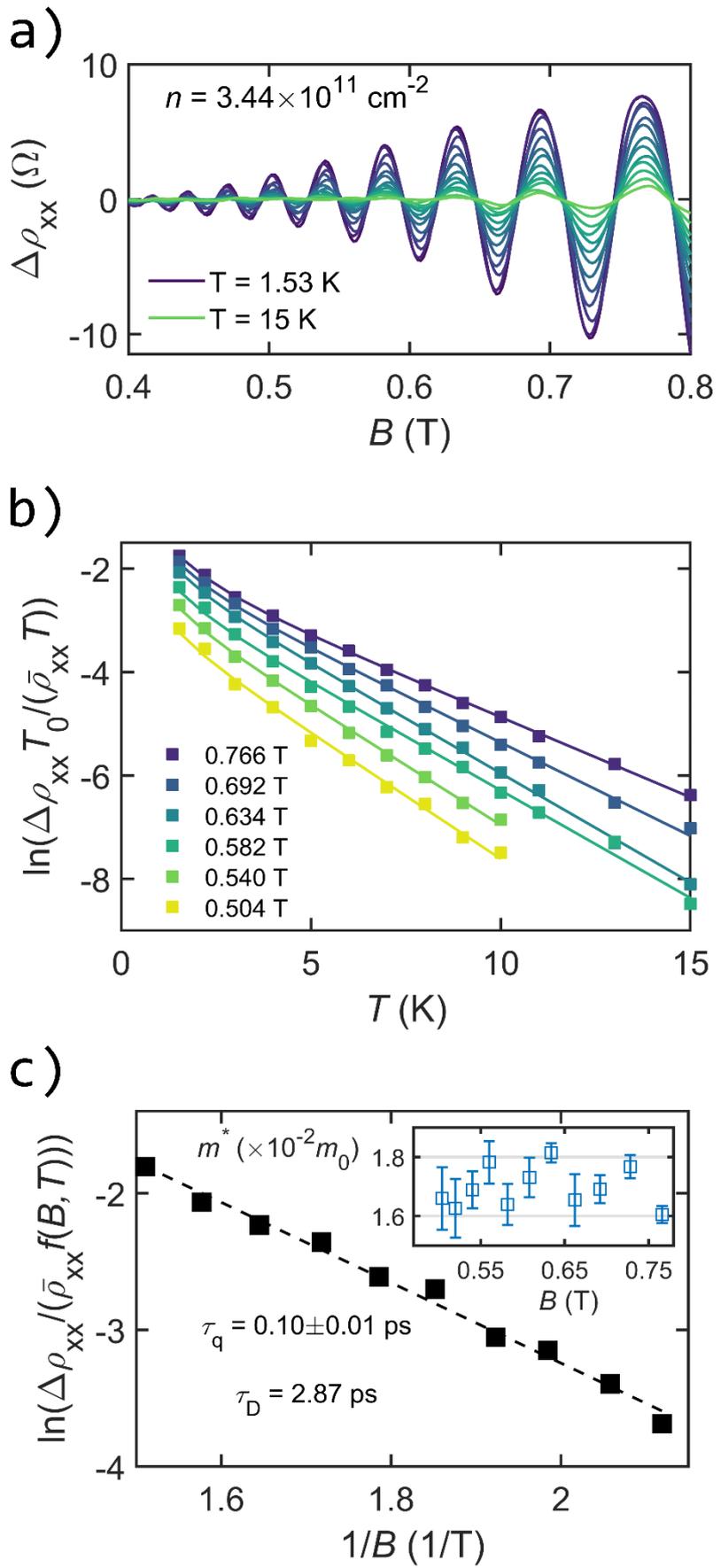

**Figure 5 (*This is a 1-column figure*)**

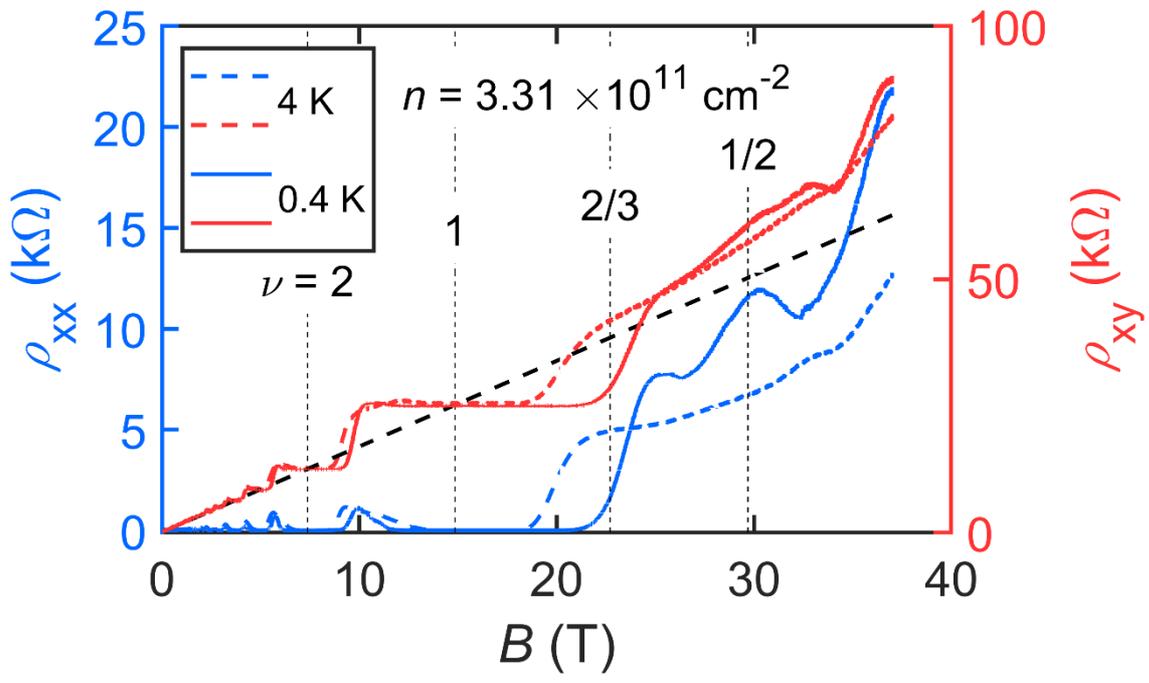